\documentclass[9pt,conference]{IEEEtran}

\usepackage[preprint]{waspaa25}

\usepackage{bm} %
\usepackage{cite}
\usepackage{amsmath,amssymb,amsfonts}
\usepackage{algorithmic}
\usepackage{graphicx}
\usepackage{textcomp}
\usepackage{xcolor}
\def\BibTeX{{\rm B\kern-.05em{\sc i\kern-.025em b}\kern-.08em
    T\kern-.1667em\lower.7ex\hbox{E}\kern-.125emX}}

\usepackage{amsmath,graphicx}
\usepackage{amsmath,amsfonts}
\usepackage{algorithmic}
\usepackage{algorithm}
\usepackage{array}
\usepackage{hyperref}
\usepackage[caption=false,font=normalsize,labelfont=sf,textfont=sf]{subfig}
\usepackage{textcomp}
\usepackage{stfloats}
\usepackage{url}
\usepackage{verbatim}
\usepackage{graphicx}
\usepackage{cite}
\usepackage{bm}
\usepackage{multirow}
\usepackage{amssymb}
\usepackage{comment}
\usepackage{mathtools}
\usepackage{booktabs}
\usepackage{balance}
\usepackage{flushend}
\usepackage{siunitx}
\usepackage{xcolor}
\usepackage{arydshln}
\usepackage{enumitem}
\AtBeginDocument{
  \abovedisplayskip     =1\abovedisplayskip
  \abovedisplayshortskip=1\abovedisplayshortskip
  \belowdisplayskip     =1\belowdisplayskip
  \belowdisplayshortskip=1\belowdisplayshortskip
}

\def\L{{\mathcal L}}

\def\C{{\mathbb C}}
\def\R{{\mathbb R}}

\title{Is MixIT Really Unsuitable for Correlated Sources?\\ Exploring MixIT for Unsupervised Pre-training in Music Source Separation}

\name{Kohei Saijo$^{1,2}$,
      Yoshiaki Bando$^{1}$}
\address{$^{1}$National Institute of Advanced Industrial Science and Technology (AIST), Tokyo, Japan \\
$^{2}$Waseda University, Tokyo, Japan
}

\begin{document}

\maketitle

\begin{abstract}
In music source separation (MSS), obtaining isolated sources or stems is highly costly, making pre-training on unlabeled data a promising approach.
Although source-agnostic unsupervised learning like mixture-invariant training (MixIT) has been explored in general sound separation, they have been largely overlooked in MSS due to its implicit assumption of source independence.
We hypothesize, however, that the difficulty of applying MixIT to MSS arises from the ill-posed nature of MSS itself, where stem definitions are application-dependent and models lack explicit knowledge of what should or should not be separated, rather than from high inter-source correlation.
While MixIT does not assume any source model and struggles with such ambiguities, our preliminary experiments show that it can still separate instruments to some extent, suggesting its potential for unsupervised pre-training.
Motivated by these insights, this study investigates MixIT-based pre-training for MSS.
We first pre-train a model on in-the-wild, unlabeled data from the Free Music Archive using MixIT, and then fine-tune it on MUSDB18 with supervision.
Using the band-split TF-Locoformer, one of the state-of-the-art MSS models, we demonstrate that MixIT-based pre-training improves the performance over training from scratch.
\end{abstract}

\section{Introduction}
\label{sec:intro}

Music source separation (MSS) has seen remarkable progress in recent years, largely driven by advances in neural networks~\cite{open_unmix, spleeter, demucs, bsrnn, bsroformer}.
These models are typically trained using supervised learning with large collections of mixture-source pairs.
However, acquiring isolated sources (or stems) for MSS is extremely costly, and publicly available datasets are scarce.
For example, the most widely used dataset in the community, MUSDB18~\cite{MUSDB18HQ}, contains only 150 tracks, approximately 10 hours for each of the vocals, drums, bass, and other (VDBO) stem.

Given the data scarcity in MSS, recent studies have explored leveraging unlabeled data.
In~\cite{pac_hubert}, HuBERT-style pre-training~\cite{hubert} for MSS, called Pac-HuBERT, has been proposed.
It uses cluster labels derived from primitive auditory features such as repetition structure~\cite{rafii2012repeating} and melody contour~\cite{salamon2012melody}.
By attaching a decoder to the pre-trained Pac-HuBERT encoder and fine-tuning with labeled data, the model outperformed training from scratch.
However, its performance heavily depends on the choice of auditory features.
In~\cite{pac_hubert}, the selected primitives were closely tied to vocals and the "other" stem, leading to notable improvements in their separation.
Still, little improvement has been reported for bass and drums in low-resource settings.

\begin{figure}[t]
\centering
\vspace{0mm}
\centerline{\includegraphics[width=\linewidth]{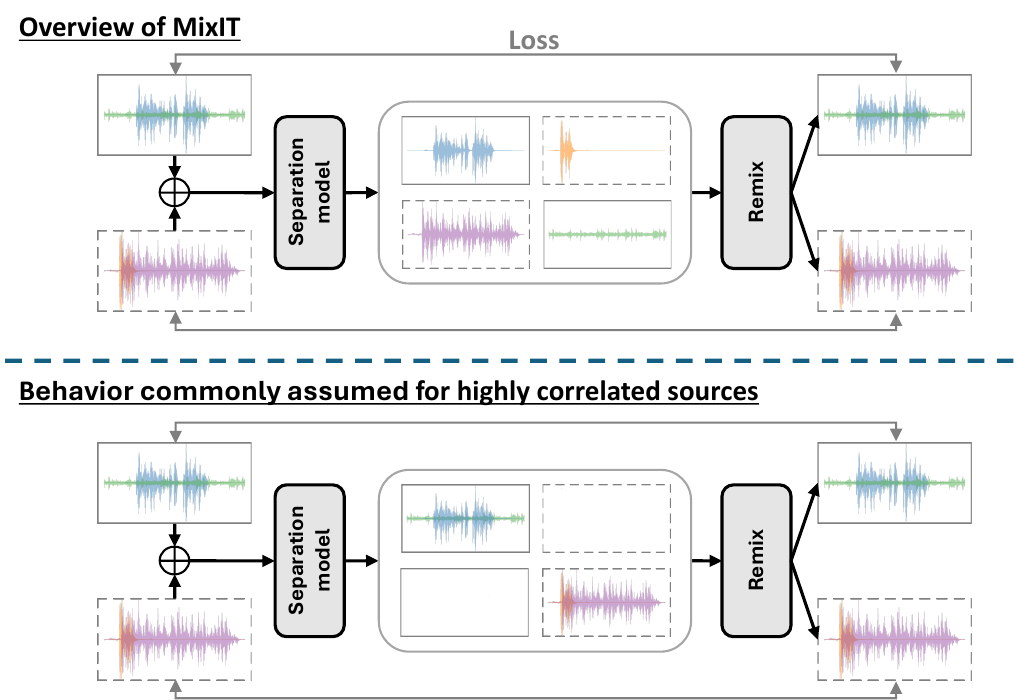}}
\caption{
   \textbf{Top}: Overview of MixIT. The model is trained to separate sources in an MoM with a mixture reconstruction loss. The model needs to separate each source well since it does not know which source originally belonged to which mixture.
   \textbf{Bottom}: Behavior of MixIT commonly assumed when inter-source correlation is high. In this case, the model may identify the combinations of the sources by finding a certain shared property among the sources in each mixture, leading to a sub-optimal solution to output the individual mixtures directly. Although this is the common assumption in MSS, we revisit MixIT and investigate its effectiveness for MSS pre-training.
}
\label{fig:mixit}
\end{figure}

In general sound separation, several source-agnostic unsupervised methods have been proposed~\cite{saijo_rccl, remixit_journal, selfremixing}.
A representative method is mixture-invariant training (MixIT)~\cite{mixit}, where the model is trained to separate a mixture of mixtures (MoM), created by adding multiple mixtures together.
The model is trained to reconstruct the individual mixtures in the MoM by remixing the separated sources with the optimal permutation (Fig.~\ref{fig:mixit}).
Since the model does not know which mixture each source in the MoM originally belonged to, it needs to separate the sources to reconstruct the original mixtures well.
Although MixIT has been shown to be effective for speech and environmental sound separation, it has been argued that MixIT does not perform well in MSS~\cite{manilow2022source, schulze2023unsupervised}.
The rationale is that high inter-source correlation in musical mixtures may allow the model to associate sources with specific mixtures, leading the training to fall into a sub-optimal solution where the model separates mixtures directly rather than disentangling individual sources (Fig.~\ref{fig:mixit} bottom).
However, such claims are based on qualitative properties, and no quantitative analysis has been reported.

We argue that it is premature to conclude without any experimental validation that MixIT is ineffective for MSS solely due to the high inter-source correlation.
This is because a non-negligible portion of data may exhibit relatively low inter-source correlation, making it feasible to learn separation from such data.
We hypothesize that MixIT's limited performance in MSS arises from the ill-posed nature of MSS itself, rather than from high inter-source correlation.
In MSS, the definition of each stem is often application-dependent, and the model lacks explicit knowledge of what should or should not be separated.
For example, the "other" stem sometimes consists of multiple instruments in the commonly considered VDBO setup, and separating it into multiple sources is penalized by standard intrusive metrics.
Consequently, models trained with MixIT may appear to perform poorly, even if they can separate sources to some extent.
Indeed, in our preliminary experiments, we have found that the model trained with MixIT on music data somewhat separates sources, although the signal-to-distortion ratio (SDR) remains low.
Although MixIT does not assume any source model and struggles with the ambiguities of the stem definition, these preliminary results suggest its potential for pre-training.

Taking these considerations into account, this study investigates the effectiveness of the MixIT for pre-training MSS models.
We first pre-train the model on a large-scale, unlabeled music dataset and then fine-tune it using labeled data.
In the experiments, we demonstrate that fine-tuning the pre-training model yields notable performance gains over training from scratch.
In addition, by fine-tuning only the encoder and decoder of an encoder-separator-decoder model, we show that the MixIT-pretrained model already achieves good performance.
We also examine the impact of pre-training data scale.
We will publish the source code and model weights once the paper is accepted.

\section{Related work}
\label{sec:related_work}

\subsection{Semi-supervised learning in music source separation}
\label{ssec:semisup_mss}
One prevailing approach to address the data scarcity is semi-supervised learning, where a model pre-trained on labeled data is used to generate pseudo-labels through source activity detection or separation~\cite{demucs, wang2021semi, bsrnn}.
Although these methods have shown promise, their effectiveness may heavily depend on the quality of the unlabeled data.
Prior work on MSS has relied on in-house collections~\cite{demucs,bsrnn}, but we do not have access to such high-quality unlabeled data.
While semi-supervised learning can potentially be combined with the unsupervised pre-training, we leave it for future work and focus on pre-training.

\subsection{Unsupervised music source separation}
\label{ssec:unsup_mss}
Neri et al. trained a variational autoencoder (VAE) by maximizing the evidence lower bound (ELBO)~\cite{neri2021unsupervised}.
In~\cite{schulze2023unsupervised}, parametric source model-based MSS is trained with a mixture reconstruction objective~\cite{schulze2023unsupervised}.
Despite their promising results, they inherently constrain the model to be based on a VAE or a parametric source model, while recent state-of-the-art MSS models are discriminative models.
For the purpose of pre-training, we are interested in a more universal model-agnostic method.
Although the method in~\cite{seetharaman2019bootstrapping}, which uses pseudo-labels obtained by an ensemble of primitive separation algorithms, is model-agnostic, it can work only for vocal extraction.

The applicability of MixIT to MSS remains underexplored, as it has often been considered unsuitable for MSS.
The original paper presents a successful example of music separation using a model trained on a large-scale dataset, primarily composed of speech and environmental sounds (Fig.6 in~\cite{mixit}).
However, this result alone does not sufficiently demonstrate that MixIT works when trained solely on music data, for two reasons: (i) only a single example is shown, and (ii) the training data is mostly non-musical.
In a follow-up work~\cite{efficient_mixit}, MixIT-trained models were evaluated on several tasks, but MSS was not included.
To the best of our knowledge, \cite{neri2021unsupervised} is the only study that has applied MixIT to MSS, but they employ random mixing to create mixtures, making the setting unrealistic.
In contrast, this study explores the effectiveness of MixIT using real in-the-wild music mixtures.

\subsection{Other source-agnostic unsupervised separation methods}
\label{ssec:unsup_ss}
A known issue in MixIT is \textit{over-separation}, where the model tends to produce more sources than necessary during inference due to training on MoMs.
Several studies have proposed techniques to mitigate this, including auxiliary losses~\cite{efficient_mixit}, teacher-student learning~\cite{ tsmixit,remixit_journal}, and additional unsupervised fine-tuning~\cite{selfremixing, selfremixing_scratch}.
However, since MixIT is used for pre-training in this study, we expect that over-separation will be mitigated in the subsequent fine-tuning stage.
Therefore, we do not explore these methods here and leave them for future work.

\section{MixIT as a pre-training method for MSS}
\label{sec:methods}

\subsection{Mixture invariant training (MixIT)}
\label{ssec:mixit}

Let us denote a mixture as $\bm{x}\in\mathbb{R}^{M \times L}$, where $M$ denotes the number of channels and $L$ denotes the number of samples in the time domain.
Since we focus on MSS, the mixtures are stereo ($M=2$).
To simplify the notation, we omit $M$ hereafter, without loss of generality.
We define a separation model with $N$ output channels as $f$.

MixIT trains a separation model $f$ in an unsupervised manner using the MoM $\bar{\bm{x}} = \bm{x}_{1} + \bm{x}_{2} \in \R^{L}$, made by mixing two different mixtures $\bm{x}_{1}$ and $\bm{x}_{2}$.
The model first separates the MoM into each source:
\begin{align}
  \label{eq:mixit_separation}
     \hat{\bm{s}} = f(\bar{\bm{x}}) \in \R^{N \times L}.
\end{align}
The MixIT loss is then computed between the separated signals $\hat{\bm{s}}$ and the individual mixtures, as described in~\cite{mixit}:
\begin{align}
  \label{eq:mixit_loss}
    \L_{\rm{MixIT}} = \min_{\bm{A}} \sum\nolimits_{b=1}^{2} {\L(\bm{x}_{b}, [\bm{A}\hat{\bm{s}}]_{b})},
\end{align}
where $\L$ is a signal-level loss function and the mixing matrix $\bm{A}\in\mathbb{B}^{{2}\times{N}}$ assigns each $\hat{s}_{n}$ to the original mixtures.

While Eq.~(\ref{eq:mixit_loss}) needs exhaustive $\mathcal{O}(2^{N})$ search to obtain the optimal mixing matrix $\bar{\bm{A}}$, we use an efficient solver proposed in~\cite{efficient_mixit}:
\begin{align}
  \label{eq:efficient_mixit_loss}
     \bar{\bm{A}} = \mathcal{P}_{\mathbb{B}}\{{\mathrm{argmin}_{\bm{A} \in \R^{2 \times N}}} ||\bm{x} - \bm{A}\hat{\bm{s}}||^{2}_{2}\},
\end{align}
where $\mathcal{P}_{\mathbb{B}}$ projects the maximum value in each column to 1 and the rest of the elements to 0 to make $\bar{\bm{A}}$ binary.
Using Eq.~(\ref{eq:efficient_mixit_loss}) greatly reduces the computational cost, while giving similar performance~\cite{efficient_mixit}.

\subsection{Discussion: Is MixIT really unsuitable for MSS?}
\label{ssec:discussion}
As described in the previous section, MixIT trains the model to separate sources such that, when remixed with the optimal permutation, they reconstruct the original mixtures well.
While this training scheme may appear to allow a sub-optimal solution, where the model directly outputs the original mixtures (e.g., $\bm{x}_1$ and $\bm{x}_2$) instead of separating them into individual sources $\bm{s}_1, \dots, \bm{s}_n$ (see Fig.~\ref{fig:mixit}), the training does not converge to such a solution if the sources are independent.
This is because the model has no information about which source belongs to which original mixture, making it infeasible to directly estimate $\bm{x}_1$ and $\bm{x}_2$ from the mixture of mixtures $\bar{\bm{x}}$.
Thus, the model is forced to separate all sources to minimize the loss.

However, this assumption may not hold when the sources within each mixture are highly correlated.
If the sources share certain common features, the model might be able to infer which sources belong to which mixture, potentially converging to the sub-optimal solution mentioned earlier.
Based on this qualitative reasoning, previous studies have argued that MixIT is unsuitable for tasks like MSS, where inter-source correlation is typically high~\cite{manilow2022source, schulze2023unsupervised}.
However, to the best of our knowledge, no experimental analysis has been reported.

We argue that it is premature to conclude that MixIT fails due to high source correlation.
Not all training data exhibit strong inter-source correlation; a substantial portion may contain sources with relatively low correlation.
Moreover, musical compositions are diverse: while instruments may share rhythm or tonality, they typically have distinct timbres.
Given this, it is more likely that, in most cases, the model can reduce the loss more easily by somewhat separating the sources rather than directly predicting the original mixtures.

We posit that the common belief that MixIT cannot work in MSS is more likely due to the ill-posed nature of the task than to high inter-source correlation.
In MSS, stem definitions are application-dependent.
For instance, in the VDBO setup, the "other" stem may contain multiple sources.
Even in stems like drums, elements such as bass drums and hi-hats, despite having distinct timbres, must be grouped and output as a single stem.
Since MixIT does not assume any explicit source model, it cannot account for such ambiguities in stem definitions.
As a result, it may over-separate sources that should remain grouped, leading to deceptively low scores on standard instrusive metrics.
In contrast, for in-the-wild data, where the types and presence of instruments are unknown, incorporating stem semantics into MixIT is not very easy.
Therefore, in this study, we focus on MixIT as a pre-training strategy for MSS, despite its limitations in handling stem ambiguity.
Such ambiguities are resolved during the subsequent supervised fine-tuning stage.

\section{Experiments}
\label{sec:experiments}

\subsection{Datasets}
\label{ssec:datasets}

\textbf{Free Music Archive (FMA)}~\cite{fma} is used for unsupervised pre-training.
The dataset is available in four different sizes, small, medium, large, and full, containing approximately 67, 209, 890, and 8,232 hours of data, respectively.
We mainly used the large subset.
Due to the long duration of audio files, we segmented them into 10-second clips with a 5-second overlap.
For each segment, we computed the power of the audio signal in 1-second intervals and discard the segment if a segment contained more than 5 seconds of silence.
To ensure consistency in sampling rate, we removed any audio with a sampling rate below 44.1 kHz and downsampled any audio with a higher rate to 44.1 kHz.
However, note that most of the content exhibits significantly reduced energy in high-frequency bands, likely due to MP3 compression, resulting in effective bandwidths often falling short of 44.1 kHz~\cite{URGENT2025}.

\textbf{MUSDB18-HQ}~\cite{MUSDB18HQ} was used for the supervised fine-tuning.
Each song is provided as a mixture of four stems (vocals, bass, drums, and other), and the goal is to decompose the mixture into these four.
MUSDB18-HQ contains 100 training tracks and 50 test tracks, all sampled at 44.1 kHz.
We followed a widely adopted split, using 86 songs for training and 14 songs for validation.
To remove silent regions, we applied the unsupervised source activity detection method introduced in~\cite{bsrnn} to the training data. %

\subsection{Separation model}
\label{ssec:models}

We used the band-split TF-locoformer (BS-locoformer) model introduced in~\cite{tuss} as the separation model since it has shown strong MSS performance~\cite{tflocoformer_nope}.
BS-Locoformer is based on the typical encoder-separator-decoder architecture, where the encoder is the band-split module~\cite{bsrnn}, the decoder is the band-wise decoding module, and the separator consists of several TF-Locoformer blocks~\cite{tflocoformer}, respectively.

The input to the encoder is a complex spectrogram after applying the short-time Fourier transform (STFT) to the mixture waveform, $\bm{X} \in \R^{2M \times T \times F}$, with $T$ time frames and $F$ frequency bins.
2 corresponds to the real and imaginary (RI) parts.
The band-split encoder decomposes the spectrogram into $Q$ non-overlapping subband spectrograms $\bm{X}_{q} \in \C^{2M \times T \times b_q}$ $(q={1,\dots ,Q})$, where the pre-defined band-widths $b_{q}$ satisfy $\sum_{q} b_{q} = F$.
$\bm{X}_{q}$ undergoes normalization and a linear transformation, producing a feature representation $\bm{Z}_{q} \in \R^{D \times T \times 1}$.
The $Q$ features are then concatenated and result in a feature $\bm{Z} \in \R^{D \times T \times Q}$, which is processed by TF-Locoformer blocks.
The band-wise decoding module splits $\bm{Z}$ into $Q$ sub-features and decodes them to obtain subband-wise complex masks with an MLP block (see \cite{bsroformer} for more details).
The separated signals in the time domain are obtained by masking and the inverse STFT.
We used the same band-split configuration as~\cite{bsroformer} with $Q=62$ bands.

The core separation part, the TF-Locoformer block, has frequency modeling and temporal modeling sub-blocks, each based on multi-head self-attention and convolutional feed-forward networks.
Please refer to the original paper~\cite{tflocoformer} for more details.
We trained models in three different sizes: small, medium, and large.
For the medium model, we set the number of TF-Locoformer blocks $B$ to 6, the embedding dimension $D$ to 128, the hidden dimension in FFNs $C$ to 192, the kernel size in convolution $K$ to 8, the stride of convolution $S$ to 1, the number of heads in self-attention $H$ to 8, the number of groups in group normalization $G$ to 8, respectively.
We set $B=4$, $D=96$, $C=128$, $H=4$, $G=4$ in the small model and $B=9$, $D=192$, $C=256$ in the large model, while keeping the other configurations unchanged.
We removed the positional encoding following~\cite{tflocoformer_nope}.

\begin{table*}[t]
\centering
\sisetup{
detect-weight, %
mode=text, %
tight-spacing=true,
round-mode=places,
round-precision=2,
table-format=2.2,
table-number-alignment=center
}
\caption{
    cSDR and uSDR [dB] on MUSDB18-HQ test set.
    "Large" split of FMA is used for MixIT pre-training.
    Results marked in grey are not directly comparable because ($\dagger$) validation split is not explicitly mentioned, ($\ddagger$) different validation split is used, ($\lozenge$) additional 1750 in-house unlabeled data is used, and ($\star$) additional 500 in-house labelled data is used.
    Cells marked with “–” indicate that scores were not reported in the original papers.
}

\label{table:results}
\resizebox{0.8\linewidth}{!}{
\begin{tabular}{lll*{12}{S}}

\toprule

\multirow{2}{*}[-1.3ex]{\texttt{ID}} &\multirow{2}{*}[-1.3ex]{\shortstack{Model}} &\multirow{2}{*}[-1.3ex]{\shortstack{Pretrain}} &\multicolumn{2}{c}{Vocals} &\multicolumn{2}{c}{Bass} &\multicolumn{2}{c}{Drums} &\multicolumn{2}{c}{Other} &\multicolumn{2}{c}{Average} \\
\cmidrule(lr){4-5}\cmidrule(lr){6-7}\cmidrule(lr){8-9}\cmidrule(lr){10-11}\cmidrule(lr){12-13}

& & &{cSDR} &{uSDR} &{cSDR} &{uSDR} &{cSDR} &{uSDR} &{cSDR} &{uSDR} &{cSDR} &{uSDR}  \\

\midrule

&\textcolor{gray}{BSRNN$^{\dagger}$~\cite{bsrnn}} &\textcolor{gray}{-} &\textcolor{gray}{10.01} &\textcolor{gray}{10.04} &\textcolor{gray}{7.22} &\textcolor{gray}{6.80} &\textcolor{gray}{9.01} &\textcolor{gray}{8.92} &\textcolor{gray}{6.70} &\textcolor{gray}{6.01} &\textcolor{gray}{8.24} &\textcolor{gray}{7.94} \\
&\textcolor{gray}{~~~+Semi-sup. fine-tune.$^{\dagger \lozenge}$~\cite{bsrnn}} &\textcolor{gray}{-} &\textcolor{gray}{10.47} &\textcolor{gray}{10.47} &\textcolor{gray}{8.16} &\textcolor{gray}{7.20} &\textcolor{gray}{10.15} &\textcolor{gray}{9.66} &\textcolor{gray}{7.08} &\textcolor{gray}{6.33} &\textcolor{gray}{8.97} &\textcolor{gray}{8.42} \\
&\textcolor{gray}{SIMO stereo BSRNN$^{\dagger}$~\cite{simo_stereo_bsrnn}} &\textcolor{gray}{-} &\textcolor{gray}{9.73} &\textcolor{gray}{10.27} &\textcolor{gray}{7.80} &\textcolor{gray}{7.61} &\textcolor{gray}{10.06} &\textcolor{gray}{9.83} &\textcolor{gray}{6.56} &\textcolor{gray}{6.50} &\textcolor{gray}{8.54} &\textcolor{gray}{8.55} \\
&\textcolor{gray}{TS-BSMAMBA2$^{\ddagger}$~\cite{mamba2_mss}} &\textcolor{gray}{-} &\textcolor{gray}{10.57} &\textcolor{gray}{10.60} &\textcolor{gray}{8.88} &\textcolor{gray}{7.47} &\textcolor{gray}{10.34} &\textcolor{gray}{10.07} &\textcolor{gray}{8.45} &\textcolor{gray}{6.69} &\textcolor{gray}{9.56} &\textcolor{gray}{8.71} \\
&\textcolor{gray}{BS-Roformer$^{\dagger}$~\cite{bsroformer}} &\textcolor{gray}{-} &\textcolor{gray}{10.66} &\textcolor{gray}{\text{-}} &\textcolor{gray}{11.31} &\textcolor{gray}{\text{-}} &\textcolor{gray}{9.49} &\textcolor{gray}{\text{-}} &\textcolor{gray}{7.73} &\textcolor{gray}{\text{-}} &\textcolor{gray}{9.80} &\textcolor{gray}{\text{-}} \\
&\textcolor{gray}{BS-Roformer (Large)$^{\dagger \star}$~\cite{bsroformer}} &\textcolor{gray}{-} &\textcolor{gray}{12.72} &\textcolor{gray}{\text{-}} &\textcolor{gray}{13.32} &\textcolor{gray}{\text{-}} &\textcolor{gray}{12.91} &\textcolor{gray}{\text{-}} &\textcolor{gray}{9.01} &\textcolor{gray}{\text{-}} &\textcolor{gray}{11.99} &\textcolor{gray}{\text{-}} \\

\midrule

&Res-U-Net~\cite{pac_hubert} &- &8.07 &\text{-} &5.78 &\text{-} &5.21 &\text{-} &5.29 &\text{-} &6.09 &\text{-}  \\
&Res-U-Net~\cite{pac_hubert} &Pac-HuBERT~\cite{pac_hubert} &8.52 &\text{-} &6.20 &\text{-} &5.76 &\text{-} &5.18 &\text{-} &6.41 &\text{-}  \\

\midrule

\texttt{S1}&BS-Locoformer~(Small) &- &9.03 &9.18 &7.99 &7.63 &9.98 &10.11 &6.59 &6.19 &8.40 &8.28  \\
\texttt{S2}&BS-Locoformer~(Small) &MixIT &9.20 &9.48 &9.00 &8.11 &10.34 &10.45 &6.66 &6.34 &8.80 &8.59  \\

\texttt{M1}&BS-Locoformer~(Medium) &- &9.66 &9.83 &9.16 &8.14 &10.24 &10.45 &7.07 &6.57 &9.04 &8.75  \\
\texttt{M2}&BS-Locoformer~(Medium) &MixIT &10.00 &10.28 &9.65 &8.87 &10.91 &11.11 &7.27 &6.93 &9.46 &9.30  \\

\texttt{L1}&BS-Locoformer~(Large) &- &10.18 &10.23 &\bfseries 10.27 &8.78 &10.65 &10.96 &7.22 &6.85 &9.58 &9.21  \\
\texttt{L2}&BS-Locoformer~(Large) &MixIT &\bfseries 10.39 &\bfseries 10.75 &10.21 &\bfseries 9.24 &\bfseries 10.96 &\bfseries 11.34 &\bfseries 8.06 &\bfseries 7.56 &\bfseries 9.90 &\bfseries9.72  \\

\bottomrule

\end{tabular}
}

\end{table*}

\begin{table}[t]
\centering
\begingroup
\setlength{\tabcolsep}{3.5pt}
\sisetup{
detect-weight, %
mode=text, %
tight-spacing=true,
round-mode=places,
round-precision=2,
table-format=2.2,
table-number-alignment=center
}
\caption{
    cSDR and uSDR[dB] on MUSDB18-HQ test set when fine-tuning only the encoder and decoder while freezing the separator.
}

\label{table:pretraining}
\resizebox{\linewidth}{!}{
\begin{tabular}{lc*{12}{S}}

\toprule

\multirow{2}{*}[-1.ex]{\shortstack{Model\\size}} &\multirow{2}{*}[-1.ex]{\shortstack{Pre\\train}} &\multicolumn{2}{c}{Vocals} &\multicolumn{2}{c}{Bass} &\multicolumn{2}{c}{Drums} &\multicolumn{2}{c}{Other} &\multicolumn{2}{c}{Average} \\
\cmidrule(lr){3-4}\cmidrule(lr){5-6}\cmidrule(lr){7-8}\cmidrule(lr){9-10}\cmidrule(lr){11-12}

& &{cSDR} &{uSDR} &{cSDR} &{uSDR} &{cSDR} &{uSDR} &{cSDR} &{uSDR} &{cSDR} &{uSDR}  \\

\midrule

Small & &3.28 &3.62 &3.49 &3.46 &4.12 &4.87 &2.19 &2.51 &3.27 &3.62  \\
Small &\checkmark &7.72 &7.83 &6.12 &5.78 &8.15 &8.26 &4.92 &4.81 &6.73 &6.67  \\ %
Medium & &3.39 &3.66 &3.56 &3.63 &3.89 &4.82 &2.53 &2.63 &3.34 &3.68  \\
Medium &\checkmark &7.92 &8.12 &7.38 &6.62 &8.75 &9.25 &5.38 &5.10 &7.36 &7.27  \\ %
Large & &3.81 &3.87 &3.82 &3.71 &4.26 &5.06 &2.52 &2.67 &3.61 &3.83  \\
Large &\checkmark &8.43 &8.04 &6.46 &6.44 &9.02 &9.09 &5.19 &4.88 &7.27 &7.11  \\

\bottomrule

\end{tabular}
}
\endgroup
\end{table}

\begin{table}[t]
\centering
\begingroup
\setlength{\tabcolsep}{3.5pt}
\sisetup{
detect-weight, %
mode=text, %
tight-spacing=true,
round-mode=places,
round-precision=2,
table-format=2.2,
table-number-alignment=center
}
\caption{
    cSDR and uSDR [dB] of the medium model on the MUSDB18-HQ test set when changing the pre-training data scale.
}

\label{table:data_scale}
\resizebox{\linewidth}{!}{
\begin{tabular}{l*{12}{S}}

\toprule

\multirow{2}{*}[-1.ex]{\shortstack{Data\\scale}} &\multicolumn{2}{c}{Vocals} &\multicolumn{2}{c}{Bass} &\multicolumn{2}{c}{Drums} &\multicolumn{2}{c}{Other} &\multicolumn{2}{c}{Average} \\
\cmidrule(lr){2-3}\cmidrule(lr){4-5}\cmidrule(lr){6-7}\cmidrule(lr){8-9}\cmidrule(lr){10-11}

&{cSDR} &{uSDR} &{cSDR} &{uSDR} &{cSDR} &{uSDR} &{cSDR} &{uSDR} &{cSDR} &{uSDR}  \\

\midrule

- &9.66 &9.83 &9.16 &8.14 &10.24 &10.45 &7.07 &6.57 &9.04 &8.75  \\
Small &9.77 &10.06 &9.33 &8.46 &10.71 &10.99 &7.39 &6.81 &9.30 &9.08  \\
Medium &9.79 &10.34 &9.63 &8.71 &10.77 &11.03 &7.43 &7.07 &9.40 &9.29  \\
Large &10.00 &10.28 &\bfseries 9.65 &8.87 &\bfseries 10.91 &11.11 &7.27 &6.93 &9.46 &9.30  \\
Full &\bfseries 10.08 &\bfseries 10.35 &9.64 &\bfseries 8.91 &10.81 &\bfseries 11.16 &\bfseries 7.52 &\bfseries 7.12 &\bfseries 9.51 &\bfseries 9.39  \\

\bottomrule

\end{tabular}
}
\endgroup
\end{table}

\subsection{Training and evaluation details}
\label{ssec:details}
In the pre-training stage, we trained the models for 150 epochs, with each epoch consisting of around 1000 training steps.
We used the AdamW optimizer~\cite{adamw} with a weight decay factor of 1e-2.
The peak learning rate was set to 1e-3 for the small and medium models, and 5e-4 for the large model, respectively.
The learning rate was linearly increased from 0 to the peak learning rate over the first 5000 training steps, and subsequently decayed by a factor of 0.965 at each epoch end.
The input was 6 seconds long, and the batch size was 128.
Gradient clipping was applied with a maximum gradient $L_2$-norm of 5.
We employed the automatic mixed precision and flash attention~\cite{dao2022flashattention}.
The negative thresholded signal-to-noise ratio (SNR) between the ground truth $y$ and the estimated signal $\hat{y}$ was used as the loss function:
\begin{align}
  \label{eq:snr_loss}
    \L_\mathrm{SNR}(y, \hat{y}) = -10\log_{10}{\frac{||y||^2}{||y-\hat{y}||^2 + \tau||y||^2}},
\end{align}
where $\tau = 10^{-3}$ is a soft threshold that clamps the SNR at 30~dB.
We set the number of output channels $N$ to 12.

In the supervised fine-tuning stage, we trained the models for 900 epochs, with each epoch consisting of 110 training steps.
All the modules in the model are fine-tuned unless otherwise stated.
The optimizer and learning-rate scheduling schemes were the same as pre-training stage, except that the learning rate was kept constant for the first 550 epochs and then decayed by a factor of 0.98 every two epochs.
The input was 6 seconds long and the batch size was 32.
During training, dynamic mixing was performed at each training step, where a segment from each stem was randomly selected, RMS-normalized, scaled by a gain uniformly sampled from [-10, 10] dB, and mixed.
Like~\cite{bsrnn}, each segment was dropped with a probability of 0.05 before mixing to simulate the mixture where the target source is inactive.
We used the negative thresholded SNR loss but modifed it to accept zero signals~\cite{fuss}.
During inference, each song was segmented into multiple 12-second chunks with a 6-second overlap, separated individually, and reconstructed using overlap-add to obtain song-level separation results.
We used the song-level SNR (uSDR)~\cite{usdr} and chunk-wise SDR (cSDR)~\cite{csdr} as the evaluation metrics.
All other configurations remained consistent with those in the pre-training stage.

Note that we need to select 4 output channels out of 12 used in the pre-training stage before fine-tuning.
To remove the redundant channels, we select 4 output channels in the following manner:
\begin{enumerate}[left=0pt]
    \item We segment each song in the validation set of the MUSDB dataset into 6-second chunks.
    
    \item We separate each chunk with the pre-trained model, and obtain the optimal permutation that maximizes the SNR between the separated outputs and the ground truths. This allows us to identify which output channel is most likely to output which instruments.
    
    \item After processing all chunks, for each instrument, we select the output channel that was most frequently aligned with that instrument across all permutations.
\end{enumerate}

\subsection{Main results}
\label{ssec:main_results}
We first investigate the effectiveness of the pre-training by comparing the fine-tuned models and those trained from scratch.
The "large" subset of the FMA is used for pre-training.
The results are presented in Table~\ref{table:results}.
For reference, Table~\ref{table:results} also includes the performance of previous state-of-the-art methods.
In addition, Table~\ref{table:results} includes the results of Pac-HuBERT, a method that performs HuBERT-style pre-training using the FMA large subset.
Methods that do not follow the commonly adapted train/validation split are colored in gray.

From Table~\ref{table:results}, we observe that fine-tuning the pretrained model consistently yields better performance across all model sizes.
Notably, for the medium and large models, uSDR improves by more than 0.5 dB (\texttt{M1} vs. \texttt{M2}, \texttt{L1} vs. \texttt{L2}).
Remarkably, pre-training enables the model to achieve performance comparable to that of a larger model trained from scratch (\texttt{S2} vs. \texttt{M1}, \texttt{M2} vs. \texttt{L1}).
In particular, while \texttt{L1} already achieves performance comparable to the state-of-the-art, \texttt{M2} reaches a similar level.
These findings highlight the value of pre-training on large-scale in-the-wild data in MSS, where labeled data are limited.
While the comparison with Pac-HuBERT is not entirely fair due to differences in model architecture, the MixIT pre-training achieves comparable or greater cSDR improvements, demonstrating its effectiveness.

\subsection{Performance of pre-trained model}
\label{ssec:pretraing_results}
We then investigate how well the pre-trained model itself performs on MSS.
Note that, as mentioned earlier, the model was not trained to follow the source definitions used in MUSDB (i.e., the VDBO setup).
Thus, instead of directly evaluating the pretrained model, we fine-tune the encoder and decoder for 200 epochs on MUSDB while keeping the core separator frozen.
This allows the model to adapt its output to match the VDBO setup, enabling us to assess the separation ability of the pre-trained separator more fairly.
The results are presented in Table~\ref{table:pretraining}.
For comparison, we also performed the same fine-tuning procedure on a randomly initialized model. %

As shown in Table~\ref{table:pretraining}, the model pre-trained with MixIT achieves significantly better separation performance than the model without pre-training.
Even when compared with the previous state-of-the-art methods, the pre-trained model works reasonably well.
For example, the medium model achieves a uSDR of 7.27~dB, which is only 0.67~dB behind the BSRNN’s 7.94~dB.
The results support our hypothesis that MixIT can work in MSS.

\subsection{Impact of data scale in pre-training}
\label{ssec:data_scale}
To investigate the impact of data size in pre-training, we trained medium-sized models using the small, medium, and full subsets of the FMA dataset.
The performances of these models after the fine-tuning are shown in Table~\ref{table:data_scale}.
All the experimental setups, other than the data scale, including the number of training steps, were kept consistent with the experiments in Section~\ref{ssec:main_results}.

As shown in Table~\ref{table:data_scale}, increasing the size of the pre-training dataset leads to improved performance.
However, the performance gains were moderate; even when using the full subset, which is nearly ten times larger than the large subset, the improvement is limited to approximately 0.1~dB.
This limited gain is likely due to the relatively small parameter size of the separator.
It is known that larger models tend to benefit more from larger datasets, but in our case, the separator in the medium BS-Locoformer contains only 15.0 million parameters. %
We plan to conduct experiments with alternative models that have larger parameter sizes in future work.

\section{Conclusion}
\label{sec:conclusion}
In this work, we investigated the effectiveness of MixIT for MSS.
Although MixIT has been considered unsuitable for tasks like MSS, where source signals are highly correlated, we hypothesized that its apparent limitations stem mainly from ambiguities in source definitions within the MSS task itself.
Motivated by this, we explored the potential of MixIT as a pre-training method for MSS.
We first pre-trained a model on the FMA dataset using MixIT, then conducted supervised fine-tuning on the MUSDB dataset.
We showed that full fine-tuning of the pre-trained model yields notable improvements over training from scratch.
By fine-tuning only the encoder and decoder while freezing the separator in a pre-trained encoder–separator–decoder model, we also demonstrated that the MixIT-pretrained model successfully learns MSS.
Additionally, we found that using larger pre-training datasets generally leads to better performance after fine-tuning.

\section{Acknowledgment}
\label{sec:ack}
This study was partially supported by JST FOREST JPMJFR232Y.

\clearpage
\bibliographystyle{IEEEtran}
\bibliography{main}

\end{document}